\documentclass[12pt,a4paper]{article}
\makeatletter
\def\eqnarray{%
\stepcounter{equation}%
\let\@currentlabel=\theequation
\global\@eqnswtrue
\global\@eqcnt\z@
\tabskip\@centering
\let\\=\@eqncr
$$\halign to \displaywidth\bgroup\@eqnsel\hskip\@centering
$\displaystyle\tabskip\z@{##}$&\global\@eqcnt\@ne
\hfil$\displaystyle{{}##{}}$\hfil
&\global\@eqcnt\tw@$\displaystyle\tabskip\z@{##}$\hfil
\tabskip\@centering&\llap{##}\tabskip\z@\cr}
\makeatother
\newcommand{\ket}[1]{{\vert{#1}\rangle}}

\newcommand{\real}{{\mathbf R}}

\begin{document}

\title{\sl Jaynes--Cummings Model and a Non--Commutative ``Geometry" : 
A Few Problems Noted}
\author{
  Kazuyuki FUJII
  \thanks{E-mail address : fujii@yokohama-cu.ac.jp}\\
  Department of Mathematical Sciences\\
  Yokohama City University\\
  Yokohama, 236--0027\\
  Japan
  }
\date{}
\maketitle
%\thispagestyle{empty}
%
%
%  gaiyou
%
%
\begin{abstract}
  In this paper we point out that the Jaynes--Cummings model without taking 
  a renonance conditon gives a non--commutative version of the simple spin 
  model (including the parameters $x$, $y$ and $z$) treated by M. V. Berry. 
  This model is different from usual non--commutative ones because the x--y 
  coordinates are quantized, while the z coordinate is not. 
  
  One of new and interesting points in our non--commutative model is that the 
  strings corresponding to Dirac ones in the Berry model exist only in states 
  containing the ground state (${\cal F}\times \{\ket{0}\} \cup 
  \{\ket{0}\}\times {\cal F}$), while for other excited states 
  (${\cal F}\times {\cal F} \setminus 
  {\cal F}\times \{\ket{0}\} \cup \{\ket{0}\}\times {\cal F}$) 
  they don't exist. 
  
  It is probable that a non--commutative model makes singular objects 
  (singular points or singular lines or etc) in the corresponding classical 
  model mild or removes them partly. 
\end{abstract}

\newpage

%
%
%     Honbun
%
%

\section{Introduction}

The Hopf bundles (which are famous examples of fiber bundles) over 
${\bf K}={\bf R}$, ${\bf C}$, ${\bf H}$ (the field of quaternion numbers), 
${\bf O}$ (the field of octanion numbers) are classical objects and 
they are never written down in a local manner. If we write them locally then 
we are forced to encounter singular lines called the Dirac strings, 
see \cite{MN}. 

It is very interesting to comment that the Hopf bundles correspond to 
topological solitons called Kink, Monopole, Instanton, Generalized Instanton 
respectively, see for example \cite{MN}, \cite{Ra}, \cite{KFu}. 
Therefore they are very important objects to study in detail. 

Berry has given the Hopf bundle and Dirac strings by making use of a 
Hamiltonian (a simple spin model including the parameters $x$, $y$ and $z$), 
see the paper in \cite{SW}. 
We call this the Berry model for simplicity in the following. 

We would like to make the Hopf bundles non--commutative. Whether such a 
generalization is meaningful or not is not clear at the current time, however 
it may be worth trying, see for example \cite{BaI} or more recently 
\cite{SG} and its references.

By the way, we are studying a quantum computation based on cavity QED and 
one of the basic tools is the Jaynes--Cummings model (or more generally 
the Tavis--Cummings one), \cite{JC}, \cite{MS}, \cite{papers}, \cite{FHKW}. 
This is given as a ``half" of the Dicke model under the resonance condition 
and rotating wave approximation associated to it. 
If the resonance condition is not taken, then this model gives a 
non--commutative version of the Berry model. However, this new one is 
different from usual one because $x$ and $y$ coordinates are quantized, 
while $z$ coordinate is not. 

If we study the non--commutative Berry model by making use of so--called 
Quantum Diagonalization Method (QDM) developed in \cite{FHKSW}, then we see 
that the Dirac strings appear in only states containing the ground one 
(${\cal F}\times \{\ket{0}\} \cup \{\ket{0}\}\times {\cal F}$) where 
${\cal F}$ is the Fock space generated by $\{a,\ a^{\dagger}\}$, 
while in excited states (${\cal F}\times {\cal F} 
\setminus {\cal F}\times \{\ket{0}\} \cup \{\ket{0}\}\times {\cal F}$) 
they don't appear. 
That is, this means that classical singularities are not universal in the 
process of non--commutativization, which is a very interesting phenomenon. 

Why do we consider non--commutative versions of classical field models ? 
What is an advantage to consider such a generalization ? 
Researchers in this subject should answer such natural questions. 
This note may give one of answers.

\section{Berry Model and Dirac Strings : Review}

First of all we explain the Dirac strings and Hopf bundle which 
Berry constructed in \cite{SW}. 
The Hamiltonian considered by Berry is a simple spin model 
\begin{equation}
\label{eq:berry-hamiltonian}
H_{B}
=x\sigma_{1}+y\sigma_{2}+z\sigma_{3}
=(x-iy)\sigma_{+}+(x+iy)\sigma_{-}+z\sigma_{3}
=
\left(
  \begin{array}{cc}
    z    & x-iy \\
    x+iy & -z
  \end{array}
\right)
\end{equation}
where $\sigma_{j}\ (j=1\sim 3)$ is the Pauli matrices, 
$\sigma_{\pm}\equiv (1/2)(\sigma_{1}\pm i\sigma_{2})$ and $x$, $y$ and $z$ 
are parameters. We would like to diagonalize $H$ above. The eigenvalues are 
\[
\lambda=\pm r\equiv \pm\sqrt{x^{2}+y^{2}+z^{2}}
\]
and corresponding orthonormal eigenvectors are 
\[
\ket{r}=\frac{1}{\sqrt{2r(r+z)}}
 \left(
  \begin{array}{c}
    r+z  \\
    x+iy   
  \end{array}
\right),\quad 
\ket{-r}=\frac{1}{\sqrt{2r(r+z)}}
 \left(
  \begin{array}{c}
    -x+iy  \\
    r+z      
  \end{array}
\right).
\]
Here we assume $(x,y,z) \in \real^{3}-\{(0,0,0)\}\equiv 
\real^{3}\setminus \{0\}$ to avoid a degenerate case. 
Therefore a unitary matrix defined by 
\begin{equation}
U_{I}=(\ket{r},\ket{-r})
=\frac{1}{\sqrt{2r(r+z)}}
\left(
  \begin{array}{cc}
    r+z  & -x+iy \\
    x+iy & r+z
  \end{array}
\right)
\end{equation}
makes $H_{B}$ diagonal like 
\begin{equation}
H_{B}=
U_{I}
\left(
  \begin{array}{cc}
    r &     \\
      & -r
  \end{array}
\right)
U_{I}^{\dagger}\equiv
U_{I}D_{B}U_{I}^{\dagger}.
\end{equation}
We note that the unitary matrix $U_{I}$ is not defined on the whole space 
$\real^{3}\setminus \{0\}$. The defining region of $U_{I}$ is 
\begin{equation}
D_{I}=\real^{3}\setminus \{0\}- \{(0,0,z)\in \real^{3}|\ z < 0\}.
\end{equation}
The removed line $\{(0,0,z)\in \real^{3}|\ z < 0\}$ 
is just the (lower) Dirac string, which is impossible to add to $D_{I}$.

Next, we have another diagonal form of $H_{B}$ like
\begin{equation}
H_{B}=U_{II}D_{B}U_{II}^{\dagger}
\end{equation}
with the unitary matrix $U_{II}$ defined by 
\begin{equation}
U_{II}
=\frac{1}{\sqrt{2r(r-z)}}
\left(
  \begin{array}{cc}
    x-iy & -r+z \\
    r-z  & x+iy
  \end{array}
\right).
\end{equation}
The defining region of $U_{I}$ is 
\begin{equation}
D_{II}=\real^{3}\setminus \{0\}- \{(0,0,z)\in \real^{3}|\ z > 0\}.
\end{equation}
The removed line $\{(0,0,z)\in \real^{3}|\ z > 0\}$ 
is just the (upper) Dirac string, which is also impossible to add to $D_{II}$.

%\vspace{-10mm}
%%%%%%%%%%%%%%%%%%%%%%%%%%%%%%%%%%%%%%%%%%%%%%%%%%%%%%%%%%%%%%
%Picture of Dirac String
\begin{figure}
\begin{center}
\setlength{\unitlength}{1mm} 
\begin{picture}(140,60)(5,0)
\put(40,20){\circle{2}}
\put(20,20){\line(1,0){19}}
\put(41,20){\vector(1,0){19}}
\put(40,21){\vector(0,1){19}}
\multiput(40,19)(0,-2.5){8}{\line(0,-1){2}}
\put(40.4,20.4){\line(1,1){15}}
\put(39.7,19.7){\vector(-1,-1){15}}
\put(100,20){\circle{2}}
\put(80,20){\line(1,0){19}}
\put(101,20){\vector(1,0){19}}
\multiput(100,21)(0,2.5){7}{\line(0,1){2}}
\put(100,37.5){\vector(0,1){2}}
\put(100,19){\line(0,-1){19}}
\put(100.4,20.4){\line(1,1){15}}
\put(99.7,19.7){\vector(-1,-1){15}}
\end{picture}
\end{center}
\caption{Dirac strings corresponding to I and II}
\end{figure}
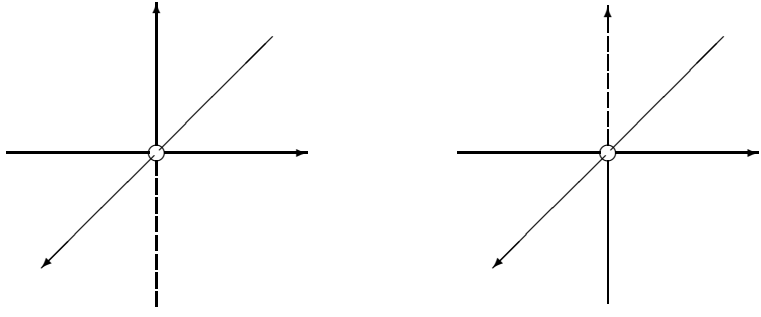
%%%%%%%%%%%%%%%%%%%%%%%%%%%%%%%%%%%%%%%%%%%%%%%%%%%%%%%%%%%%%%%%

Here we have diagonalizations of two types for $H$, so a natural question 
comes about. What is a relation between $U_{I}$ and $U_{II}$ ? 
If we define 
\begin{equation}
\Phi
=\frac{1}{\sqrt{x^{2}+y^{2}}}
\left(
  \begin{array}{cc}
    x-iy &      \\
         & x+iy
  \end{array}
\right)
\end{equation}
then it is easy to see 
\[
U_{II}=U_{I}\Phi.
\]
We note that $\Phi$ (which is called a transition function) is not defined on 
the whole $z$--axis. 

What we would like to emphasize here is that the diagonalization of $H$ is 
not given globally (on $\real^{3}\setminus \{0\}$). However, the dynamics is 
perfectly controlled by the system 
\begin{equation}
\left\{(U_{I},D_{I}), (U_{II},D_{II}), \Phi, \real^{3}\setminus \{0\}=
D_{I}\cup D_{II}\right\},
\end{equation}
which defines a famous fiber bundle called the Hopf bundle associated to the 
complex numbers ${\bf C}$ \footnote{The base space $\real^{3}\setminus \{0\}$ 
is homotopic to the two--dimensional sphere $S^{2}$},
\[
S^{1}\longrightarrow S^{3}\longrightarrow S^{2},
\]
see \cite{MN}.

\section{All Hopf Bundles and Dirac Strings}

In this section we show that the contents in the preceding section are 
easily generalized to the all Hopf bundles ($n=1, 2, 4, 8$). Then as a 
by-product Dirac strings associated to them are shown clearly. 

Let ${\bf K}$ be the field of real numbers ${\bf R}$, of complex numbers 
${\bf C}$, of quaternion numbers ${\bf H}$, 
of octanion numbers ${\bf O}$ respectively. 
We write an element of ${\bf K}$ by 
\[
w=\sum_{j=0}^{n-1}x_{j}{\bf k}_{j},\quad x_{j} \in {\bf R},
\]
where ${\bf k}_{j}$ are generators of ${\bf K}$. Explicitly 
\begin{eqnarray}
w&=&x_{0}\quad \mbox{for}\quad {\bf K}={\bf R}, \\
w&=&x_{0}+x_{1}i\quad \mbox{for}\quad {\bf K}={\bf C}, \\
w&=&x_{0}+x_{1}i+x_{2}j+x_{3}k\quad \mbox{for}\quad {\bf K}={\bf H}, \\
w&=&x_{0}+\sum_{j=1}^{7}x_{j}e_{j}\quad \mbox{for}\quad {\bf K}={\bf O}
\end{eqnarray}
and $\bar{w}$ is a conjugate of $w$ in ${\bf K}$. Then it is well--known that 
$
\bar{w}w=w\bar{w}=||w||^{2}=\sum_{j=0}^{n-1}x_{j}^{2}. 
$
We note that $n=\mbox{dim}_{{\bf R}}{\bf K}=1,2,4,8$ respectively. 

As a ``unified" Hamiltonian whose model space is ${\bf K}\times {\bf R}$ 
we consider 
\begin{equation}
\label{eq:general-hamiltonian}
H_{{\bf K}}=w\sigma_{+}+\bar{w}\sigma_{-}+z\sigma_{3}
=
\left(
  \begin{array}{cc}
    z & \bar{w}  \\
    w & -z
  \end{array}
\right)
\end{equation}
where $z\in {\bf R}$. Of course, $H_{{\bf C}}$ = $H_{B}$ in 
(\ref{eq:berry-hamiltonian}).

Then we have a decomposition of $H_{\bf K}$ like 
\begin{equation}
\label{eq:diagonal-decomposition}
H_{{\bf K}}=
\left\{
\begin{array}{l}
U_{I}D_{{\bf K}}U_{I}^{\dagger}\quad on\quad D_{I}  \\
U_{II}D_{{\bf K}}U_{II}^{\dagger}\quad on\quad D_{II}  
\end{array}
\right.
\end{equation}
where 
\begin{equation}
U_{I}
=\frac{1}{\sqrt{2r(r+z)}}
\left(
  \begin{array}{cc}
    r+z & -\bar{w} \\
    w & r+z
  \end{array}
\right),\quad
U_{II}
=\frac{1}{\sqrt{2r(r-z)}}
\left(
  \begin{array}{cc}
    \bar{w} & -r+z \\
    r-z  & w
  \end{array}
\right)
\end{equation}
for $r=\sqrt{||w||^{2}+z^{2}}$ and 
\[
D_{{\bf K}}=
\left(
  \begin{array}{cc}
    r &     \\
      & -r
  \end{array}
\right),\quad 
\Phi_{{\bf K}}=U_{I}^{\dagger}U_{II}=
\frac{1}{||w||}
\left(
  \begin{array}{cc}
    \bar{w} &   \\
            & w 
  \end{array}
\right).
\]
Since we are not interested in the degenerate case, we assume that 
$r\ne 0$ in the following ($(w,z)\in {\bf K}\times {\bf R}\setminus (0,0)
\equiv {\bf K}\times {\bf R}\setminus \{0\}$ \footnote{${\bf K}\times {\bf R}
\setminus \{0\}$ is homotopic to the $n$--dimensional sphere $S^{n}$}).

What we would like to emphasize here is that the diagonalization of 
$H_{{\bf K}}$ is not given globally, so there are Dirac strings. However, 
the dynamics is perfectly controlled by the system 
\begin{equation}
\left\{(U_{I},D_{I}), (U_{II},D_{II}), \Phi_{{\bf K}}, 
{\bf K}\times {\bf R}\setminus \{0\}=D_{I}\cup D_{II}\right\},
\end{equation}
which defines famous fiber bundles called the Hopf bundles
\begin{eqnarray}
{\bf Z_{2}}&\longrightarrow& S^{1}\longrightarrow S^{1} 
\quad \mbox{for}\quad {\bf K}={\bf R}, \\
U(1)&\longrightarrow& S^{3}\longrightarrow S^{2} 
\quad \mbox{for}\quad {\bf K}={\bf C}, \\
Sp(1)&\longrightarrow& S^{7}\longrightarrow S^{4} 
\quad \mbox{for}\quad {\bf K}={\bf H}, \\
S{\bf o}(1)&\longrightarrow& S^{15}\longrightarrow S^{8} 
\quad \mbox{for}\quad {\bf K}={\bf O}, 
\end{eqnarray}
where $U(1)\cong S^{1}$, $Sp(1)\cong S^{3}$ and $S{\bf o}(1)\cong S^{7}$ are 
well--known, \cite{MN}. 

The projectors corresponding to the Hopf bundles are given as 
\begin{equation}
\label{eq:projectors}
P(w,z)=U_{I}P_{0}U_{I}^{\dagger}=U_{II}P_{0}U_{II}^{\dagger}
=
\frac{1}{2r}
\left(
  \begin{array}{cc}
    r+z & \bar{w}  \\
     w  & r-z
  \end{array}
\right),\quad r=\sqrt{||w||^{2}+z^{2}}
\end{equation}
where $P_{0}$ is a basic one 
\[
P_{0}=
\left(
  \begin{array}{cc}
    1 &   \\
      & 0
  \end{array}
\right)\ \ {\in}
 \ \ M(2,{\bf K}).
\]
We note that in (\ref{eq:projectors}) Dirac strings don't appear because 
the projectors $P$ are expressed globally.

It is interesting to note that the Hopf bundles correspond to topological 
solitons called Kink, Monopole, Instanton, Generalized Instanton 
respectively, see for example \cite{Ra}, \cite{KFu}.

\section{Two Steps Decomposition}

With the decomposition (\ref{eq:diagonal-decomposition}) it is not easy to see 
where the Dirac strings come from. To see this point we give in this section 
two steps decomposition to the Hamiltonian (\ref{eq:general-hamiltonian}), 
which makes the Dirac strings of Hopf bundles clear. It is easy to see 
\begin{equation}
\label{eq:unusual-decomposition}
H_{{\bf K}}
=
\left(
  \begin{array}{cc}
    z & \bar{w}  \\
    w & -z
  \end{array}
\right)
=
\left(
  \begin{array}{cc}
    1 &                 \\
      & \frac{w}{||w||}
  \end{array}
\right)
\left(
  \begin{array}{cc}
    z    & ||w||  \\
   ||w|| & -z
  \end{array}
\right)
\left(
  \begin{array}{cc}
    1 &                        \\
      & \frac{\bar{w}}{||w||}
  \end{array}
\right),
\end{equation}
so the middle matrix 
\[
\left(
  \begin{array}{cc}
    z    & ||w||  \\
   ||w|| & -z
  \end{array}
\right)
\]
which is common to all ${\bf R}$, ${\bf C}$, ${\bf H}$ and ${\bf O}$, 
play a central role in the Dirac strings. Now let us diagonalize it to be 
\begin{equation}
\label{eq:dirac-strings}
\left(
  \begin{array}{cc}
    z    & ||w||  \\
   ||w|| & -z
  \end{array}
\right)
=
\left\{
\begin{array}{l}
 U_{I}D_{{\bf K}}U_{I}^{\dagger}   \\
 U_{II}D_{{\bf K}}U_{II}^{\dagger} 
\end{array}
\right.
\end{equation}
where $r=\sqrt{||w||^{2}+z^{2}}$ and 
\[
U_{I}=
 \frac{1}{\sqrt{2r(r+z)}}
 \left(
  \begin{array}{cc}
    r+z  & -||w|| \\
   ||w|| & r+z
  \end{array}
\right),
\quad 
U_{II}=
 \frac{1}{\sqrt{2r(r-z)}}
 \left(
  \begin{array}{cc}
    ||w|| & -r+z    \\
     r-z  & ||w||
  \end{array}
\right).
\]
We in this stage encounter the Dirac strings. 
It is just this matrix (\ref{eq:dirac-strings}) that bears the Dirac strings 
on its shoulders.

\section{A Non--Commutative Berry Model from the Jaynes--Cummings Model}

First, let us explain the Jaynes--Cummings model which is well--known in 
quantum optics, \cite{JC}. 
The Hamiltonian of Jaynes--Cummings model can be written as follows (we set 
$\hbar=1$ for simplicity) 
\begin{equation}
\label{eq:hamiltonian}
H=
\omega 1_{2}\otimes a^{\dagger}a + \frac{\Delta}{2} \sigma_{3}\otimes {\bf 1} 
+ g\left(\sigma_{+}\otimes a+\sigma_{-}\otimes a^{\dagger} \right),
\end{equation}
where $\omega$ is the frequency of single radiation field, $\Delta$ the energy 
difference of two level atom, $a$ and $a^{\dagger}$ are annihilation and 
creation operators of the field, and $g$ a coupling constant. We assume that 
$g$ is small enough (a weak coupling regime). 
Here $\sigma_{+}$, $\sigma_{-}$ and $\sigma_{3}$ are given as 
\begin{equation}
\label{eq:sigmas}
\sigma_{+}=
\left(
  \begin{array}{cc}
    0& 1 \\
    0& 0
  \end{array}
\right), \quad 
\sigma_{-}=
\left(
  \begin{array}{cc}
    0& 0 \\
    1& 0
  \end{array}
\right), \quad 
\sigma_{3}=
\left(
  \begin{array}{cc}
    1& 0  \\
    0& -1
  \end{array}
\right), \quad 
1_{2}=
\left(
  \begin{array}{cc}
    1& 0  \\
    0& 1
  \end{array}
\right).
\end{equation}
See the figure 2 as an image of the Jaynes--Cummings model. 

%%%%%%%%%%%%%%%%%%%%%%%%%%%%%%%%%%%%%%%%%%%%%%%%%%%%%%%%%%
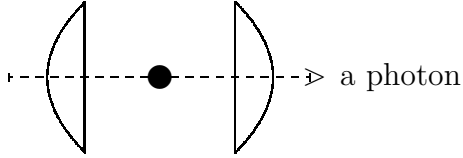
\begin{figure}
\begin{center}
\setlength{\unitlength}{1mm} 
\begin{picture}(80,40)(0,-20)
\bezier{200}(20,0)(10,10)(20,20)
\put(20,0){\line(0,1){20}}
\put(30,10){\circle*{3}}
\bezier{200}(40,0)(50,10)(40,20)
\put(40,0){\line(0,1){20}}
\put(10,10){\dashbox(40,0)}
\put(49,9){$>$}
\put(54,9){a photon}
\end{picture}
\end{center}
\vspace{-20mm}
\caption{One atom and a single photon inserted in a cavity}
\end{figure}
%%%%%%%%%%%%%%%%%%%%%%%%%%%%%%%%%%%%%%%%%%%%%%%%%%%%%%%%%%

Now we consider the evolution operator of the model. 
We rewrite the Hamiltonian (\ref{eq:hamiltonian}) as follows. 
\begin{equation}
\label{eq:hamiltonian-rewrite}
H=
\omega 1_{2}\otimes a^{\dagger}a + \frac{\omega}{2}\sigma_{3}\otimes {\bf 1} + 
\frac{\Delta-\omega}{2} \sigma_{3}\otimes {\bf 1} +
g\left(\sigma_{+}\otimes a+\sigma_{-}\otimes a^{\dagger} \right)
\equiv H_{1}+H_{2}.
\end{equation}
Then it is easy to see $[H_{1},H_{2}]=0$, which leads to 
$
\mbox{e}^{-itH}=\mbox{e}^{-itH_{1}}\mbox{e}^{-itH_{2}}.
$

In the following we consider $\mbox{e}^{-itH_{2}}$ in which 
the resonance condition $\Delta-\omega=0$ is not taken. For simplicity 
we set $\theta=\frac{\Delta-\omega}{2g}(\neq 0)$ \footnote{Since the 
Jaynes--Cummings model is obtained by the Dicke model under some resonance 
condition on parameters included, it is nothing but an approximate one 
in the neighborhood of the point, so we must assume that $|\theta|$ is small 
enough} then 
\[
%\begin{equation}
H_{2}
=g\left(
\sigma_{+}\otimes a+\sigma_{-}\otimes a^{\dagger} + 
\frac{\Delta-\omega}{2g}\sigma_{3}\otimes {\bf 1}
\right)
=g\left(
\sigma_{+}\otimes a+\sigma_{-}\otimes a^{\dagger} + 
\theta \sigma_{3}\otimes {\bf 1}
\right).
%\end{equation}
\]
For further simplicity we set 
\begin{equation}
\label{eq:non-commutative Berry model}
H_{JC}
=\sigma_{+}\otimes a+\sigma_{-}\otimes a^{\dagger} + 
\theta \sigma_{3}\otimes {\bf 1}
=
\left(
  \begin{array}{cc}
    \theta      & a       \\
    a^{\dagger} & -\theta
  \end{array}
\right),\quad [a,a^{\dagger}]={\bf 1}
\end{equation}
where we have written $\theta$ in place of $\theta {\bf 1}$ for simplicity. 

$H_{JC}$ can be considered as a non-commutative version of $H_{B}$ under the 
correspondence 
$
a\ \longleftrightarrow\ x-iy,\ a^{\dagger}\ \longleftrightarrow\ x+iy \ 
\mbox{and}\ \theta\ \longleftrightarrow\ z
$. 
That is, $x$ and $y$ coordinates are quantized, while $z$ coordinate is not, 
which is different from usual one, see for example \cite{BaI}. 
It may be possible for us to call this {\bf a non--commutative Berry model}. 
We note that this model is derived not ``by hand" but by the model in quantum 
optics itself. 

We usually analyze (\ref{eq:non-commutative Berry model}) by reducing it to 
each component contained in $H(2,{\bf C})$, which is a typical analytic 
method. However, we don't adopt such a method. That is, we treat 
(\ref{eq:non-commutative Berry model}) as a kind of bundle, which means a 
``geometric" method in the title. 

To study Dirac strings in this quantized model let us decompose 
the Hamiltonian (\ref{eq:non-commutative Berry model}) like in Section 4. 
It is easy to see 
\begin{equation}
\label{eq:non-commutative decomposition}
H_{JC}
=
\left(
  \begin{array}{cc}
    \theta      & a       \\
    a^{\dagger} & -\theta
  \end{array}
\right)
=
\left(
  \begin{array}{cc}
    1 &                                 \\
      & a^{\dagger}\frac{1}{\sqrt{N+1}}
  \end{array}
\right)
\left(
  \begin{array}{cc}
     \theta   & \sqrt{N+1} \\
   \sqrt{N+1} & -\theta
  \end{array}
\right)
\left(
  \begin{array}{cc}
    1 &                       \\
      & \frac{1}{\sqrt{N+1}}a
  \end{array}
\right)
\end{equation}
from (\ref{eq:unusual-decomposition}) and \cite{FHKSW}, where $N$ is 
the number operator $N=a^{\dagger}a$. Then the middle matrix 
in the right hand side can be considered as a classical one, so we can 
diagonalize it by making use of (\ref{eq:dirac-strings}) 

\begin{equation}
\label{eq:middle matrix decomposition}
\left(
  \begin{array}{cc}
     \theta   & \sqrt{N+1} \\
   \sqrt{N+1} & -\theta
  \end{array}
\right)
=
\left\{
\begin{array}{l}
 U_{I}
 \left(
  \begin{array}{cc}
    R(N+1) &         \\
           & -R(N+1)
  \end{array}
 \right)
 U_{I}^{\dagger}\\
 U_{II}
 \left(
  \begin{array}{cc}
    R(N+1) &         \\
           & -R(N+1)
  \end{array}
 \right)
 U_{II}^{\dagger} 
\end{array}
\right.
\end{equation}
where $R(N)=\sqrt{N+\theta^{2}}$ and $U_{I}$, $U_{II}$ are defined by 
\begin{eqnarray}
\label{eq:unitary-i}
U_{I}&=&
\frac{1}{\sqrt{2R(N+1)(R(N+1)+\theta)}}
\left(
  \begin{array}{cc}
   R(N+1)+\theta & -\sqrt{N+1}   \\
   \sqrt{N+1}    & R(N+1)+\theta
  \end{array}
\right),  \\
\label{eq:unitary-ii}
U_{II}&=&
\frac{1}{\sqrt{2R(N+1)(R(N+1)-\theta)}}
\left(
  \begin{array}{cc}
      \sqrt{N+1}   & -R(N+1)+\theta \\
     R(N+1)-\theta & \sqrt{N+1} 
  \end{array}
\right). 
\end{eqnarray}
Now let us rewrite (\ref{eq:non-commutative decomposition}) by making use of 
(\ref{eq:middle matrix decomposition}) with (\ref{eq:unitary-i}). 
Inserting the identity 
\[
\left(
  \begin{array}{cc}
    1 &                       \\
      & \frac{1}{\sqrt{N+1}}a
  \end{array}
\right)
\left(
  \begin{array}{cc}
    1 &                                 \\
      & a^{\dagger}\frac{1}{\sqrt{N+1}}
  \end{array}
\right)
=
\left(
  \begin{array}{cc}
    1 &   \\
      & 1
  \end{array}
\right)
\]
gives 
\begin{eqnarray}
H_{JC}=
&&\left(
  \begin{array}{cc}
    1 &                                 \\
      & a^{\dagger}\frac{1}{\sqrt{N+1}}
  \end{array}
\right)
U_{I}
\left(
  \begin{array}{cc}
    R(N+1) &         \\
           & -R(N+1)
  \end{array}
\right)
U_{I}^{\dagger}
\left(
  \begin{array}{cc}
    1 &                       \\
      & \frac{1}{\sqrt{N+1}}a
  \end{array}
\right)    \nonumber    \\
=
&&
\left(
  \begin{array}{cc}
    1 &                                 \\
      & a^{\dagger}\frac{1}{\sqrt{N+1}}
  \end{array}
\right)
U_{I}
\left(
  \begin{array}{cc}
    1 &                       \\
      & \frac{1}{\sqrt{N+1}}a
  \end{array}
\right)
\left(
  \begin{array}{cc}
    1 &                                 \\
      & a^{\dagger}\frac{1}{\sqrt{N+1}}
  \end{array}
\right)
\left(
  \begin{array}{cc}
    R(N+1) &         \\
           & -R(N+1)
  \end{array}
\right)\times   \nonumber  \\
&&
\left(
  \begin{array}{cc}
    1 &                       \\
      & \frac{1}{\sqrt{N+1}}a
  \end{array}
\right)
\left(
  \begin{array}{cc}
    1 &                                 \\
      & a^{\dagger}\frac{1}{\sqrt{N+1}}
  \end{array}
\right)
U_{I}^{\dagger}
\left(
  \begin{array}{cc}
    1 &                       \\
      & \frac{1}{\sqrt{N+1}}a
  \end{array}
\right)         \nonumber  \\
=
&&
V_{I}
\left(
  \begin{array}{cc}
    R(N+1) &         \\
           & -R(N)
  \end{array}
\right)
V_{I}^{\dagger}, 
\end{eqnarray}
where
\begin{eqnarray*}
V_{I}
&=&
\left(
  \begin{array}{cc}
    \frac{1}{\sqrt{2R(N+1)(R(N+1)+\theta)}} &      \\
           & \frac{1}{\sqrt{2R(N)(R(N)+\theta)}}
  \end{array}
\right)
\left(
  \begin{array}{cc}
      R(N+1)+\theta & -a            \\
      a^{\dagger}   &  R(N)+\theta
  \end{array}
\right)  \\
&=&
\left(
  \begin{array}{cc}
      R(N+1)+\theta & -a            \\
      a^{\dagger}   &  R(N)+\theta
  \end{array}
\right)  
\left(
  \begin{array}{cc}
    \frac{1}{\sqrt{2R(N+1)(R(N+1)+\theta)}} &      \\
           & \frac{1}{\sqrt{2R(N)(R(N)+\theta)}}
  \end{array}
\right).
\end{eqnarray*}

Similarly, we can rewrite (\ref{eq:non-commutative decomposition}) by making 
use of (\ref{eq:middle matrix decomposition}) with (\ref{eq:unitary-ii}). 
By inserting the identity 
\[
\left(
  \begin{array}{cc}
    \frac{1}{\sqrt{N+1}}a &    \\
                          & 1
  \end{array}
\right)
\left(
  \begin{array}{cc}
    a^{\dagger}\frac{1}{\sqrt{N+1}} &    \\
                                    & 1
  \end{array}
\right)
=
\left(
  \begin{array}{cc}
    1 &   \\
      & 1
  \end{array}
\right)
\]
we obtain 
\begin{equation}
H_{JC}=
V_{II}
\left(
  \begin{array}{cc}
    R(N) &          \\
         & -R(N+1)
  \end{array}
\right)
V_{II}^{\dagger}, 
\end{equation}
where 
\begin{eqnarray*}
V_{II}
&=&
\left(
  \begin{array}{cc}
    \frac{1}{\sqrt{2R(N+1)(R(N+1)-\theta)}} &     \\
           & \frac{1}{\sqrt{2R(N)(R(N)-\theta)}}
  \end{array}
\right)
\left(
  \begin{array}{cc}
          a       & -R(N+1)+\theta  \\
      R(N)-\theta &  a^{\dagger} 
  \end{array}
\right) \\
&=&
\left(
  \begin{array}{cc}
          a       & -R(N+1)+\theta  \\
      R(N)-\theta &  a^{\dagger} 
  \end{array}
\right) 
\left(
  \begin{array}{cc}
    \frac{1}{\sqrt{2R(N)(R(N)-\theta)}} &             \\
           & \frac{1}{\sqrt{2R(N+1)(R(N+1)-\theta)}}
  \end{array}
\right).
\end{eqnarray*}

Tidying up these we have 
\begin{equation}
\label{eq:final-decomposition}
H_{JC}=
\left\{
\begin{array}{l}
V_{I}
\left(
  \begin{array}{cc}
    R(N+1) &         \\
           & -R(N)
  \end{array}
\right)
V_{I}^{\dagger}\\
V_{II}
\left(
  \begin{array}{cc}
    R(N) &         \\
         & -R(N+1)
  \end{array}
\right)
V_{II}^{\dagger}
\end{array}
\right.
\end{equation}
with $V_{I}$ and $V_{II}$ above. From the equations 
\[
R(N+1)\ket{0}=\sqrt{1+\theta^{2}}>\theta, \quad
R(N)\ket{0}=\sqrt{\theta^{2}}=|\theta|\geq \theta
\]
we know that the strings corresponding to Dirac ones exist in only states 
${\cal F}\times \ket{0} \cup \ket{0}\times {\cal F}$ where ${\cal F}$ is 
the Fock space generated by $\{a,\ a^{\dagger}\}$, while in other excited 
states ${\cal F}\times {\cal F}\setminus 
{\cal F}\times \ket{0} \cup \ket{0}\times {\cal F}$ they don't exist
\footnote{We have identified ${\cal F}\times {\cal F}$ with the space of 
$2$--component vectors over ${\cal F}$}, see the figure 3. 
The phenomenon is very interesting. 
For simplicity we again call these strings Dirac ones in the following. 
%%%%%%%%%%%%%%%%%%%%%%%%%%%%%%%%%%%%%%%%%%%%%%%%%%%%%%%%%%%%%%
%Picture of Lattice
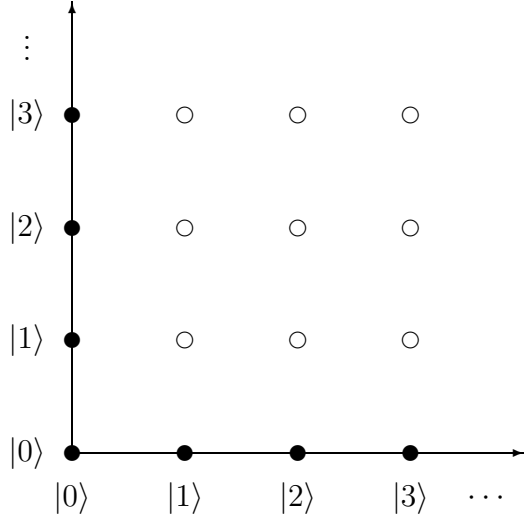
\begin{figure}
\begin{center}
\setlength{\unitlength}{1mm} 
\begin{picture}(80,70)
\put(10,10){\vector(1,0){60}}
\put(10,10){\vector(0,1){60}}
\put(10,10){\circle*{2}}
\put(25,10){\circle*{2}}
\put(40,10){\circle*{2}}
\put(55,10){\circle*{2}}
%
%\put(10,10){\circle*{2}}
\put(10,25){\circle*{2}}
\put(10,40){\circle*{2}}
\put(10,55){\circle*{2}}
\put(25,25){\circle{2}}
\put(25,40){\circle{2}}
\put(25,55){\circle{2}}
\put(40,25){\circle{2}}
\put(40,40){\circle{2}}
\put(40,55){\circle{2}}
\put(55,25){\circle{2}}
\put(55,40){\circle{2}}
\put(55,55){\circle{2}}
\put(7,1){\makebox(6,6)[c]{$|0\rangle$}}
\put(22,1){\makebox(6,6)[c]{$|1\rangle$}}
\put(37,1){\makebox(6,6)[c]{$|2\rangle$}}
\put(52,1){\makebox(6,6)[c]{$|3\rangle$}}
\put(62,1){\makebox(6,6)[c]{$\cdots$}}
\put(1,7){\makebox(6,6)[c]{$|0\rangle$}}
\put(1,22){\makebox(6,6)[c]{$|1\rangle$}}
\put(1,37){\makebox(6,6)[c]{$|2\rangle$}}
\put(1,52){\makebox(6,6)[c]{$|3\rangle$}}
\put(1,62){\makebox(6,6)[c]{$\vdots$}}
\end{picture}
\vspace{-5mm}
\caption{The bases of ${\cal F}\times {\cal F}$. The black circle means 
bases giving Dirac strings, while the white one don't.}
\end{center}
\end{figure}
%%%%%%%%%%%%%%%%%%%%%%%%%%%%%%%%%%%%%%%%%%%%%%%%%%%%%%%%%%%%%%%%%%%

Then the transition ``function" (operator) is given by 
\[
\Phi_{JC}=
\left(
  \begin{array}{cc}
    a\frac{1}{\sqrt{N}} &                               \\
                        & \frac{1}{\sqrt{N}}a^{\dagger}
  \end{array}
\right)
=
\left(
  \begin{array}{cc}
    \frac{1}{\sqrt{N+1}}a &                                 \\
                          & a^{\dagger}\frac{1}{\sqrt{N+1}}
  \end{array}
\right).
\]
The projector in this case is 
\begin{eqnarray}
\label{eq:quantum-projector}
P_{JC}&=&
V_{I}
\left(
  \begin{array}{cc}
    {\bf 1} &         \\
            & {\bf 0}
  \end{array}
\right)
V_{I}^{\dagger}
=
V_{II}
\left(
  \begin{array}{cc}
    {\bf 1} &         \\
            & {\bf 0}
  \end{array}
\right)
V_{II}^{\dagger}  \nonumber \\
&=&
\left\{
\begin{array}{l}
\left(
  \begin{array}{cc}
    \frac{1}{2R(N+1)} &                  \\
                      & \frac{1}{2R(N)} 
  \end{array}
\right)
\left(
  \begin{array}{cc}
      R(N+1)+\theta &  a            \\
      a^{\dagger}   &  R(N)-\theta
  \end{array}
\right)      \\
\left(
  \begin{array}{cc}
      R(N+1)+\theta &  a            \\
      a^{\dagger}   &  R(N)-\theta
  \end{array}
\right)
\left(
  \begin{array}{cc}
    \frac{1}{2R(N+1)} &                  \\
                      & \frac{1}{2R(N)} 
  \end{array}
\right)
\end{array}
\right.
\end{eqnarray}
, so we obtain a quantum version of (classical) spectral decomposition 
(a ``quantum spectral decomposition" by Suzuki \cite{TS}) 
\begin{equation}
\label{eq:quantum-spectral-decomposition}
H_{JC}=
\left(
  \begin{array}{cc}
    R(N+1) &       \\
           & R(N)
  \end{array}
\right)
P_{JC}
-
\left(
  \begin{array}{cc}
    R(N+1) &       \\
           & R(N)
  \end{array}
\right)
({\bf 1}_{2}-P_{JC}).
\end{equation}

As a bonus of the decomposition let us rederive the calculation of 
$\mbox{e}^{-igtH_{JC}}$ which has been given in \cite{MS}. 
The result is 
\begin{equation}
\mbox{e}^{-igtH_{JC}}=
\left(
  \begin{array}{cc}
   \mbox{cos}(tgR(N+1))-i\theta\frac{\mbox{sin}(tgR(N+1))}{R(N+1)}& 
   -i\frac{\mbox{sin}(tgR(N+1))}{R(N+1)}a       \\
   -i\frac{\mbox{sin}(tgR(N))}{R(N)}a^{\dagger}
   & \mbox{cos}(tgR(N))+i\theta\frac{\mbox{sin}(tgR(N))}{R(N)}
  \end{array}
\right)
\end{equation}
by making use of (\ref{eq:final-decomposition}) (or 
(\ref{eq:quantum-spectral-decomposition})). We leave it to the readers. 

\vspace{5mm}
Lastly in this section we make a comment on the book \cite{MS}. It is very 
interesting from a not only quantum optical but also geometric point of view. 
We believe strongly that crucial results in \cite{MS} must be reobtained from 
a ``geometric" method developed in this paper.

\section{Discussion}

In this paper we showed that a non--commutative version of the Berry model 
derived from the Jaynes--Cummings model (in quantum optics) had not Dirac 
strings in excited states. They appear in only states containing the ground 
one (${\cal F}\times \ket{0} \cup \ket{0}\times {\cal F} \subset 
{\cal F}\times {\cal F}$).

In general, a non-commutative version of classical field theory is of course 
not unique. If our model is a ``correct" one, then this paper give an example 
that classical singularities like Dirac strings are not universal in some 
non--commutative model. 
As to general case with higher spins which are not easy see \cite{TS}.

More generally, it is probable that a singularity (singularities) in some 
classical model is (are) removed in the process of non--commutativization. 
Further study on both finding many examples and constructing a general theory 
will be required.

\vspace{5mm}
\noindent{\em Acknowledgment.}\\
The author wishes to thank Akira Asada and Tatsuo Suzuki for their helpful 
comments and suggestions.

\par \vspace{5mm}
\begin{center}
 \begin{Large}
   {\bf Appendix}
 \end{Large}
\end{center}

\begin{Large}
\par \vspace{5mm} \noindent
{\bf \ \ A Local Coordinate of the Projector}
\end{Large}

In this appendix we consider a meaning of the projector 
(\ref{eq:quantum-projector}) from the view point of (infinite dimensional) 
Grassmann manifold. 
As a general introduction to this topic \cite{Fujii} is recommended. 

For that let us look for a ``local coordinate" $Z$ giving the global 
expression (\ref{eq:quantum-projector}). 
By making use of the expression by Oike in \cite{Fujii} (we don't 
repeat it here) 
\begin{equation}
\label{eq:oike-expression}
{\cal P}(Z)=
\left(
  \begin{array}{cc}
    {\bf 1} & -Z^{\dagger} \\
     Z      & {\bf 1}
  \end{array}
\right)
\left(
  \begin{array}{cc}
    {\bf 1} &          \\
            & {\bf 0}
  \end{array}
\right)
\left(
  \begin{array}{cc}
    {\bf 1} & -Z^{\dagger} \\
     Z      & {\bf 1}
  \end{array}
\right)^{-1}
\end{equation}
where $Z$ is some operator on the Fock space ${\cal F}$. Let us rewrite this 
into more useful form. From the simple relation 
\[
\left(
  \begin{array}{cc}
    {\bf 1} & Z^{\dagger} \\
     -Z     & {\bf 1}
  \end{array}
\right)
\left(
  \begin{array}{cc}
    {\bf 1} & -Z^{\dagger} \\
     Z      & {\bf 1}
  \end{array}
\right)
=
\left(
  \begin{array}{cc}
    {\bf 1}+Z^{\dagger}Z  &                      \\
                          & {\bf 1}+ZZ^{\dagger}
  \end{array}
\right)
\]
we have 
\[
\left(
  \begin{array}{cc}
    {\bf 1} & -Z^{\dagger} \\
     Z      & {\bf 1}
  \end{array}
\right)^{-1}
=
\left(
  \begin{array}{cc}
    ({\bf 1}+Z^{\dagger}Z)^{-1}  &                             \\
                                 & ({\bf 1}+ZZ^{\dagger})^{-1}
  \end{array}
\right)
\left(
  \begin{array}{cc}
    {\bf 1} & Z^{\dagger} \\
     -Z     & {\bf 1}
  \end{array}
\right).
\]
Inserting this into (\ref{eq:oike-expression}) and some calculation leads to 
\begin{equation}
\label{eq:oike-expression-2}
{\cal P}(Z)=
\left(
  \begin{array}{cc}
  ({\bf 1}+Z^{\dagger}Z)^{-1} & ({\bf 1}+Z^{\dagger}Z)^{-1}Z^{\dagger}    \\
  Z ({\bf 1}+Z^{\dagger}Z)^{-1} & Z({\bf 1}+Z^{\dagger}Z)^{-1}Z^{\dagger}
  \end{array}
\right).
\end{equation}
Comparing (\ref{eq:oike-expression-2}) with (\ref{eq:quantum-projector}) 
we finally obtain the ``local coordinate" 
\begin{equation}
\label{eq:quantum local coordinate}
Z=\frac{1}{R(N)+\theta}a^{\dagger}=a^{\dagger}\frac{1}{R(N+1)+\theta}
\end{equation}
where $R(N)=\sqrt{N+\theta^{2}}$. This is relatively simple and beautiful. 

Now if we take a classical limit $a \longrightarrow x-iy$, 
$a^{\dagger} \longrightarrow x+iy$ and $\theta=z$ then 
\begin{equation}
Z_{c}=\frac{x+iy}{r+z}
\end{equation}
where $r=\sqrt{x^{2}+y^{2}+z^{2}}$. This is nothing but a well--known one for 
(\ref{eq:projectors}) with $w=x+iy$.

%%%%%%%%%%%%%
%References%
%%%%%%%%%%%%%

\end{document}